\documentclass[9pt]{article}
\usepackage{spconf,amsmath,graphicx}

\usepackage{amssymb,amsfonts}

\usepackage{tikz,pgfplots}

\usepackage{amsthm}

\usepackage{multirow}
\usepackage{url}

\theoremstyle{plain}
\newtheorem{thm}{Theorem}[section]

\newtheorem{prop}[thm]{Proposition}

\theoremstyle{definition}

\theoremstyle{remark}

\newcommand{\xtrue}{x^\natural}
\newcommand{\xhat}{\hat{x}}
\newcommand{\abs}[1]{\left\vert #1 \right\vert}
\newcommand{\norm}[1]{\left\Vert #1 \right\Vert}
\newcommand{\set}[1]{\left\{ #1 \right\}}
\newcommand{\FF}{\mathcal{F}}

\newcommand{\Eemp}{\hat{\mathbb{E}}_m}

\newcommand{\ztrue}{z^\natural}

\title{Learning Data Triage: Linear Decoding Works for Compressive MRI}
\name{Yen-Huan~Li and Volkan~Cevher\thanks{This work was supported in part by the European Commission under Grant MIRG-268398, ERC Future Proof, SNF 200021-132548, SNF 200021-146750 and SNF CRSII2-147633.}}
\address{Laboratory for Information Inference Systems\\
\'{E}cole Polytechnique F\'{e}d\'{e}rale de Lausanne}
\begin{document}
%
\maketitle
\begin{abstract}
The standard approach to compressive sampling considers recovering an unknown deterministic signal with certain known structure, and designing the sub-sampling pattern and recovery algorithm based on the known structure. This approach requires looking for a good representation that reveals the signal structure, and solving a non-smooth convex minimization problem (e.g., basis pursuit). In this paper, another approach is considered: We learn a good sub-sampling pattern based on available training signals, without knowing the signal structure in advance, and reconstruct an accordingly sub-sampled signal by computationally much cheaper linear reconstruction. We provide a theoretical guarantee on the recovery error, and show via experiments on real-world MRI data the effectiveness of the proposed compressive MRI scheme.
\end{abstract}
%
\section{Introduction} \label{sec_intro}

The standard theory of compressive sampling (CS) considers recovering an unknown deterministic signal with certain \emph{known} structure, and designing sampling and recovery schemes based on the known structure \cite{Foucart2013}. For example, if the unknown signal is known to be sparse, one can measure it by a sub-sampling matrix satisfying the restricted isometry property (RIP), and apply basis pursuit to obtain an estimate of the signal \cite{Candes2008,Candes2006a}. Similar ideas can be extended to low-rank matrix recovery \cite{Candes2009b}, and in general signal recovery problems where the signal structures can be encoded by atomic norms or other convex functions \cite{Bach2013a,Chandrasekaran2012,ElHalabi2015}.

Despite its success in many applications, we note that there are some undesired features of the standard CS theory: 
\begin{enumerate}
\item The signal structure must be known \emph{in advance}. This usually requires seeking for a good signal representation to reveal the signal structure, a non-trivial task called dictionary learning \cite{Tosic2011}.
\item The recovery scheme is \emph{computationally expensive}. Typical examples are basis pursuit and the Lasso; both are non-smooth convex optimization problems.
\end{enumerate}


While those features seem to be necessary according to existing literature on CS, in some applications, the real-world setting can deviate from the standard setting of CS. This creates an opportunity of getting rid of those undesired features. 

We focus on one important observation which the standard CS theory does not take into consideration---we usually have training signals, i.e., signals that are given and similar to the unknown signal in some sense. 

In fact, practitioners are indeed applying this learning-based approach in a na\"{\i}ve way. For example, it is by examining a large amount of real-world images that we discovered sparsity or more sophisticated structures, under proper representations \cite{Baraniuk2010,Cevher2009,Mallat2009}. Although this na\"{\i}ve learning procedure can be made rigorous and automated by dictionary learning, training signals are still required.

In this paper, we propose alternative to compressive sampling, and apply it to compressive magnetic resonance imaging (MRI). The proposed scheme \emph{automatically} adapts to the given training signals, without any \textit{a priori} knowledge on the signal structure. We highlight the following contributions:
\begin{enumerate}
\item We propose a novel statistical learning view point to the compressive sampling problem, which allows us to study the effect of training signals.
\item Our compressive MRI scheme is computationally efficient: The learning procedure can be cast as a combinatorial optimization program, which can be exactly solved by an efficient algorithm; the recovery algorithm we consider is simply least-squares (LS) reconstruction. 
\item In contrast to the standard approach using random sub-sampling patterns \cite{Candes2006a,Lustig2007,Roman2014}, our sub-sampling scheme is fixed given the training signals, and hence simpler for implementation.
\item We provide a theoretical guarantee on the reconstruction error, and characterize its dependence on the number of training signals. 
\item We show via experiments on real MRI images that the reconstruction error performance of the proposed scheme is comparable to the performance using a finely-tuned sub-sampling pattern given in \cite{Lustig2007}. 
\end{enumerate}

\section{Review of Existing Approaches} 

Compressive MRI is essentially a linear inverse problem. The goal is to recover an unknown signal $\xtrue \in \mathbb{C}^p$, given a a sub-sampling pattern $\Omega \subset \set{ 1, \ldots, p }$ with $\abs{ \Omega } = n$ for some $n < p$, and the outcome of compressive sampling: 
\begin{equation}
y := P_{\Omega} \FF \xtrue \notag
\end{equation}
where $\FF: \mathbb{C}^p \to \mathbb{C}^p$ is the Fourier transform matrix, and $P_{\Omega}: \mathbb{C}^p \to \mathbb{C}^n$ is a linear operator that only keeps entries of $\FF \xtrue$ indexed by $\Omega$. In practice, $\xtrue$ is usually a 2D or 3D object, and $\FF$ should be replaced by the corresponding multidimensional Fourier transform. 

Existing approaches to compressive MRI can be briefly summarized as follows: 
\begin{enumerate}
\item Find a wavelet transform matrix $\Psi : \mathbb{C}^p \to \mathbb{C}^p$, such that $\xtrue = \Psi^{-1} \ztrue$ and $\ztrue$ possesses certain \emph{structure}. For example, the sparsity of $\ztrue$ and smoothness of $\xtrue$ were exploited in \cite{Lustig2007}, the tree sparsity of $\ztrue$ was considered in \cite{Chen2012a}, and the multi-level sparsity of $\ztrue$ was considered in \cite{Roman2014}.
\item Choose a \emph{random} sub-sampling pattern $\Omega$ and sample $\FF \xtrue$ accordingly; the probability distribution might be dependent on the knowledge about the structure of $\ztrue$ \cite{Lustig2007,Roman2014}. 
\item Finally, apply \emph{non-linear} decoding algorithms to reconstruct $\xtrue$. The standard basis pursuit estimator was considered in \cite{Roman2014}. A basis pursuit like estimator minimizing a linear combination of the $\ell_1$-norm and the total variation semi-norm was proposed in \cite{Lustig2007}. A closely-related Lasso like estimator with the $\ell_1$-norm and total variation semi-norm penalization was considered in \cite{Yang2010}. A similar Lasso like estimator with one additional penalization term for tree sparsity was introduced by \cite{Chen2012a}.
\end{enumerate}

We note that existing approaches essentially follow the standard theory of compressive sampling, and hence inherit the two undesired features which we mentioned in the introduction.

\section{Learning Data Triage}



The standard approach to compressive MRI models $\xtrue$ as a deterministic unknown signal. Here we adopt another modeling philosophy: We assume that $\xtrue$ is a random vector following some \emph{unknown} probability distribution $\mathbb{P}$, and we have access to $m$ training signals $x_1, \ldots, x_m \in \mathbb{C}^p$, which are independent and identically distributed random vectors also following $\mathbb{P}$, and are independent of $\xtrue$. Note that this is different from Bayesian compressive sampling \cite{Ji2008}, as $\mathbb{P}$ is unknown in our model.

We consider LS reconstruction. For any given sub-sampling pattern $\Omega$, the estimator has an explicit form: 
\begin{align}
\xhat_{\Omega} &= \arg \min_{x} \set{ \norm{ y - P_{\Omega} \FF x }_2^2 : x \in \mathbb{R}^p } \notag \\
&= \FF^H P_{\Omega}^T y. \notag
\end{align}
Once the reconstruction scheme is fixed, the only issue is to choose $\Omega$ that optimizes the resulting estimation performance.

We show in Section \ref{sec_prof_eqR} that for any given $\Omega$, the expected normalized reconstruction error satisfies
\begin{equation}
\mathbb{E}\, \frac{\norm{ \xhat_{\Omega} - \xtrue }_2^2}{\norm{ \xtrue }_2^2} = 1 - \mathbb{E}\, f_{\Omega} ( x ), \label{eq_R}
\end{equation}
where the expectations are with respect to $\xtrue \sim \mathbb{P}$ and $x \sim \mathbb{P}$, respectively, and we define
\begin{equation}
f_{\Omega} ( x ) := \frac{\norm{ P_{\Omega} \FF x }_2^2}{\norm{ x }_2^2} \notag
\end{equation}
for convenience. This implies that the optimal sub-sampling pattern $\Omega$, denoted by $\Omega_{\text{opt.}}$, is given by any solution of the following combinatorial optimization program: 
\begin{equation}
\Omega_{\text{opt}} \in \arg \max_{\Omega} \set{ \mathbb{E}\, f_{\Omega} ( x ) : \Omega \subset \set{ 1, \ldots, p }, \abs{ \Omega } = n }. \label{eq_RM}
\end{equation}
However, since $\mathbb{P}$ is assumed unknown, the optimization program is not tractable.

Motivated by the idea of empirical risk minimization in statistical learning theory \cite{Vapnik1999}, we make use of the training signals and approximate $\Omega_{\text{opt.}}$ via any solution of the optimization program: 
\begin{equation}
\Omega_m \in \arg \max_{\Omega} \set{ \hat{\mathbb{E}}_m \, f_{\Omega} ( x ) : \Omega \subset \set{ 1, \ldots, p }, \abs{ \Omega } = n } \label{eq_ERM}
\end{equation}
where $\hat{\mathbb{E}}_m$ denotes the expectation with respect to the empirical measure, i.e., 
\begin{equation}
\hat{\mathbb{E}}_m\, f_{\Omega} ( x ) := \frac{1}{m} \sum_{i = 1}^m \frac{\norm{ P_{\Omega} \FF x_i }_2^2}{\norm{ x_i }_2^2}. \notag
\end{equation}
This optimization program is tractable, because we only need to solve it for \emph{any realization of} the training signals $x_1, \ldots, x_m$. 
Note that then $\Eemp\, f_{\Omega} ( x )$ depends on $x_1, \ldots, x_m$ and is random, and so does $\Omega_m$.

The overall systems is summarized as follows: 
\begin{enumerate}
\item Find a sub-sampling pattern $\Omega_m$ by (\ref{eq_ERM}).
\item Sub-sample $\xtrue$ using $\Omega_m$ and obtain the measurement outcome 
\begin{equation}
y := P_{\Omega_m} \FF \xtrue. \notag
\end{equation}
\item Recover $\xtrue$ by 
\begin{equation}
\xhat := \FF^H P_{\Omega_m}^T y. \notag
\end{equation}
\end{enumerate}

\noindent
\textbf{On Computing $\Omega_m$:} The optimization program (\ref{eq_ERM}) is modular, and hence can be exactly solved by a simple greedy algorithm \cite{Fujishige2005}: Let $\phi_i^T$ be the $i$-th row of $\Phi$. Compute the values
\begin{equation}
v_i := \frac{1}{m} \sum_{i = 1}^m \left( \phi_i^T x_i \right)^2, \notag
\end{equation}
and set $\Omega_m$ as the set of indices corresponding to the largest $n$ $v_i$'s. The computational complexity is dominated by computation of $v_i$'s, which behaves as $\mathcal{O} ( m p ^2 )$ in general, and $\mathcal{O} ( m p \log p )$ if $\Phi$ is suitably structured, such as the Fourier and Hadamard matrices.

\section{Performance Analysis}

We analyze the reconstruction error of the proposed learning-based compressive sampling system.

If we could solve the optimization program (\ref{eq_RM}), the estimation performance would be given by
\begin{equation}
\mathbb{E}\, \frac{\norm{ \xhat_{\Omega_{\text{opt}}} - \xtrue }_2^2}{\norm{ \xtrue }_2^2} = 1 - \varepsilon_{\mathbb{P}}, \notag
\end{equation}
where we define 
\begin{equation}
\varepsilon_{\mathbb{P}} := \max_{\Omega} \set{ \mathbb{E}\, f_{\Omega} ( x ) : \Omega \subset \set{ 1, \ldots, p }, \abs{ \Omega } = n }. \notag
\end{equation}
Note that $\varepsilon_{\mathbb{P}}$ is a constant for any given $\mathbb{P}$, independent of the training signals.

Since now the optimization program (\ref{eq_RM}) is replaced by its empirical version (\ref{eq_ERM}), a reasonable guess is that the estimation performance would behave as
\begin{equation}
\mathbb{E}\, \frac{\norm{ \xhat_{\Omega_m} - \xtrue }_2^2}{\norm{ \xtrue }_2^2} \leq 1 - \varepsilon_{\mathbb{P}} + \varepsilon_m, \notag
\end{equation}
for some $\varepsilon_m > 0$ with high probability (with respect to the training signals $x_1, \ldots, x_m$), and $\varepsilon_m$ should converge to $0$ as $m \to \infty$. The following proposition verifies this guess.

\begin{prop} \label{prop_main}
For any $\beta \in ( 0, 1 )$, we have
\begin{equation}
\varepsilon_m \leq \sqrt{ \frac{2}{m} \left[ \log { p \choose n } + \log \frac{2}{\beta} \right] } \notag
\end{equation}
with probability at least $1 - \beta$.
\end{prop}

This means a size of training signals of the order $\mathcal{O} ( n \log p )$ suffices to have small enough $\varepsilon_m$, with high probability. We note that this is a worst-case guarantee, as it is \emph{distribution-independent}. In practice, $m$ can be much smaller.

\newcommand{\eqR}{\ref{eq_R}}
\newcommand{\propMain}{\ref{prop_main}}

\section{Numerical Results}
We use a 3-dimensional dataset of raw knee-images data given in $k$-space.\footnote{Available at \url{http://mridata.org/fullysampled}} We first take an inverse Fourier transform along the $z$-axis and eliminate low energy the $z-$slices that are close to the boundary of the datacube. These are noiselike slices that do not exhibit any knee feature as they are close to the skin of the patient. We then investigate subsampling schemes in the $320 \times 320$ $x-y$ Fourier plane, which corresponds to compressive sampling for each $z$-slice.

We pick the first 10 of the patients in the given dataset for training and test the learned subsampling maps on the remaining 10 patients. We compare our learning based approach to the variable density function proposed by \cite{Lustig2007}, which is parametrized by the radius of fully sampled region, $r$, and the polynomial degree, $d$. We tune the values of $r$ and $d$ so that they yield the highest average PSNR on the training data.

\begin{figure}
\centering
\begin{tabular}{cccc}
\includegraphics[width=.3\columnwidth]{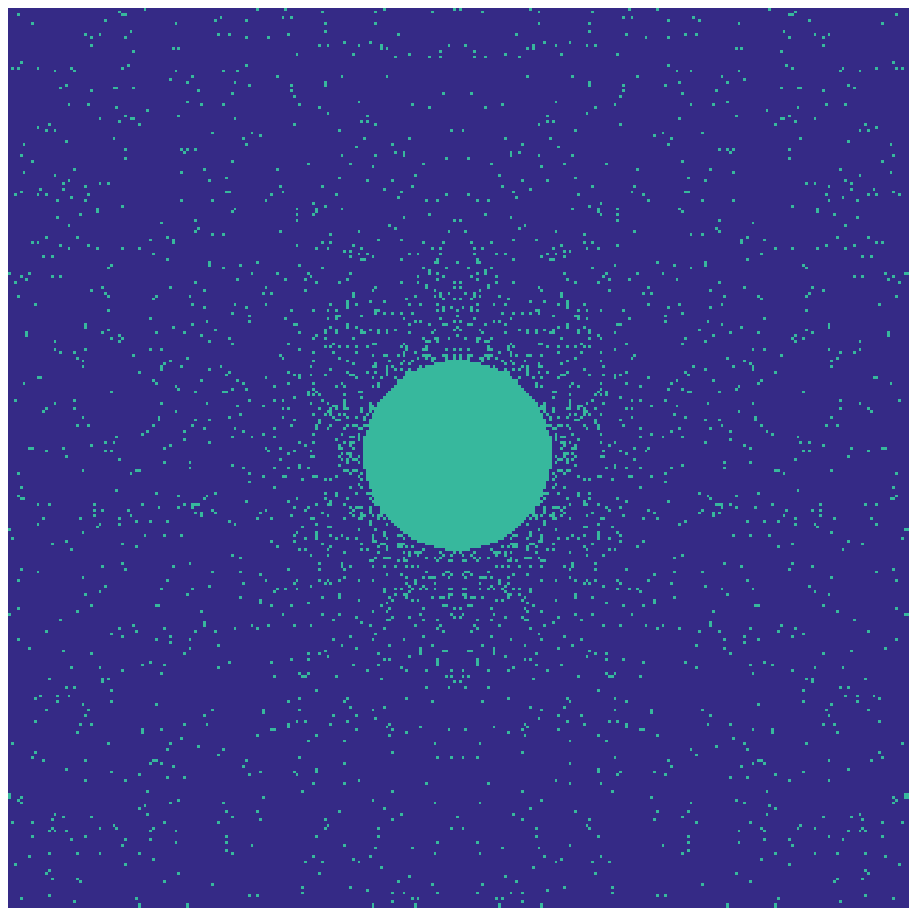} &
\hspace{-4mm}\includegraphics[width=.3\columnwidth]{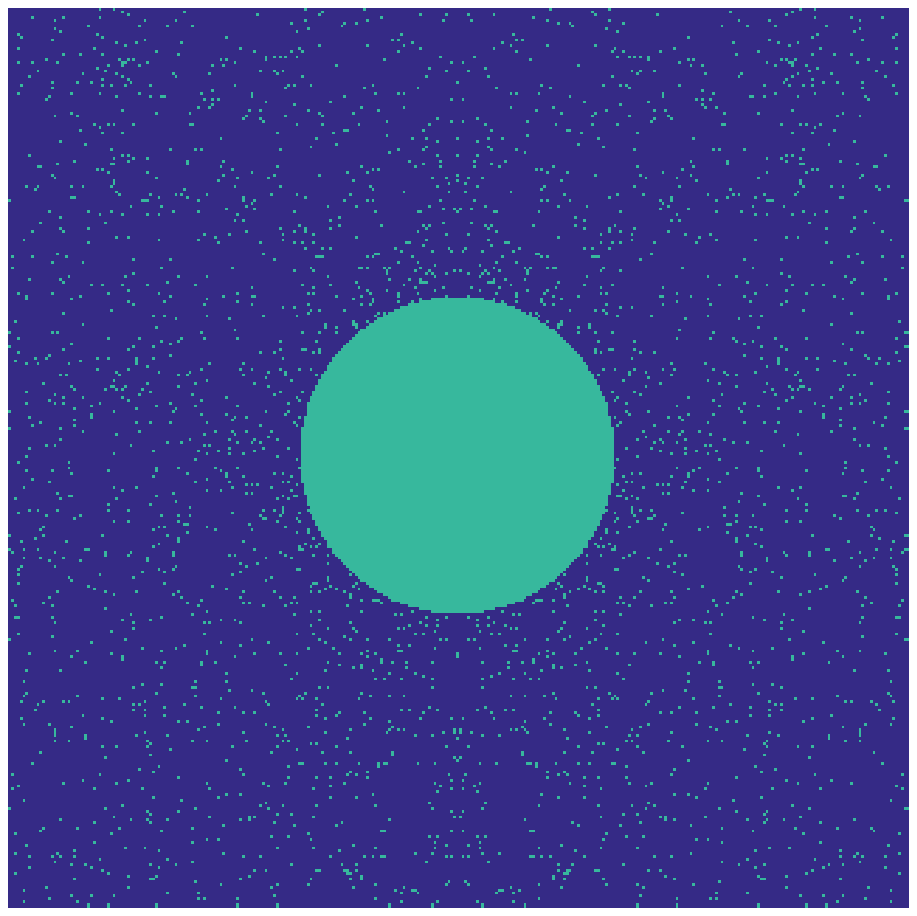} &
\hspace{-4mm}\includegraphics[width=.3\columnwidth]{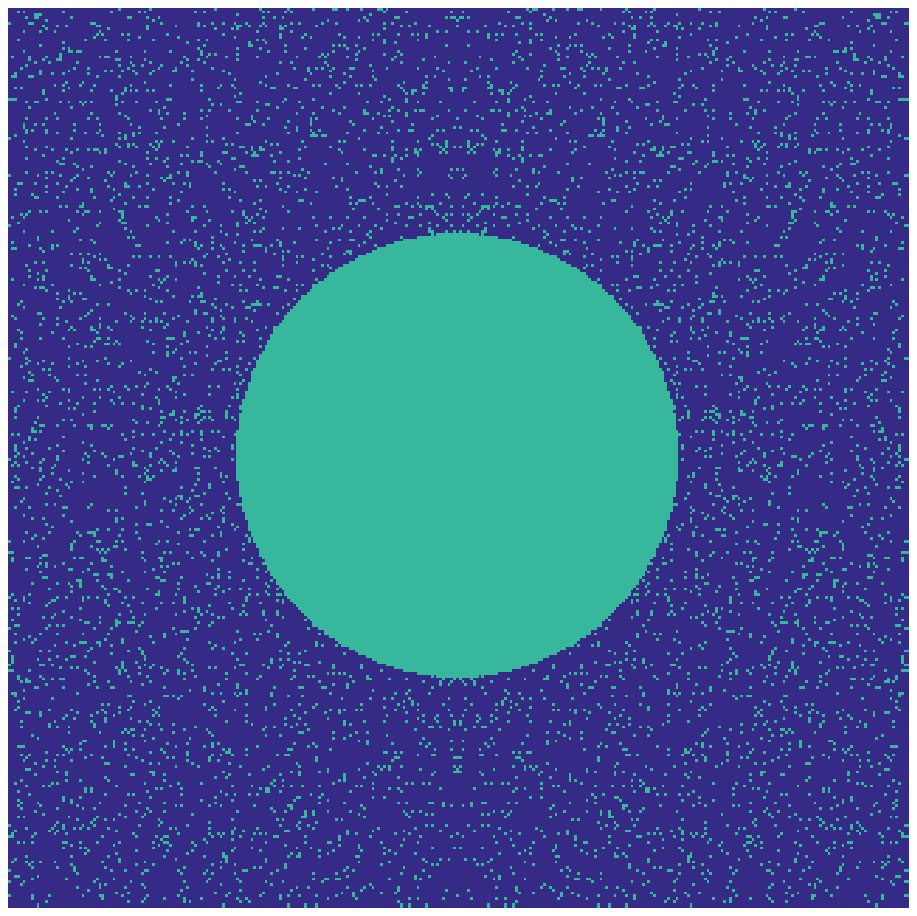}  \\[-.3mm]
\includegraphics[width=.3\columnwidth]{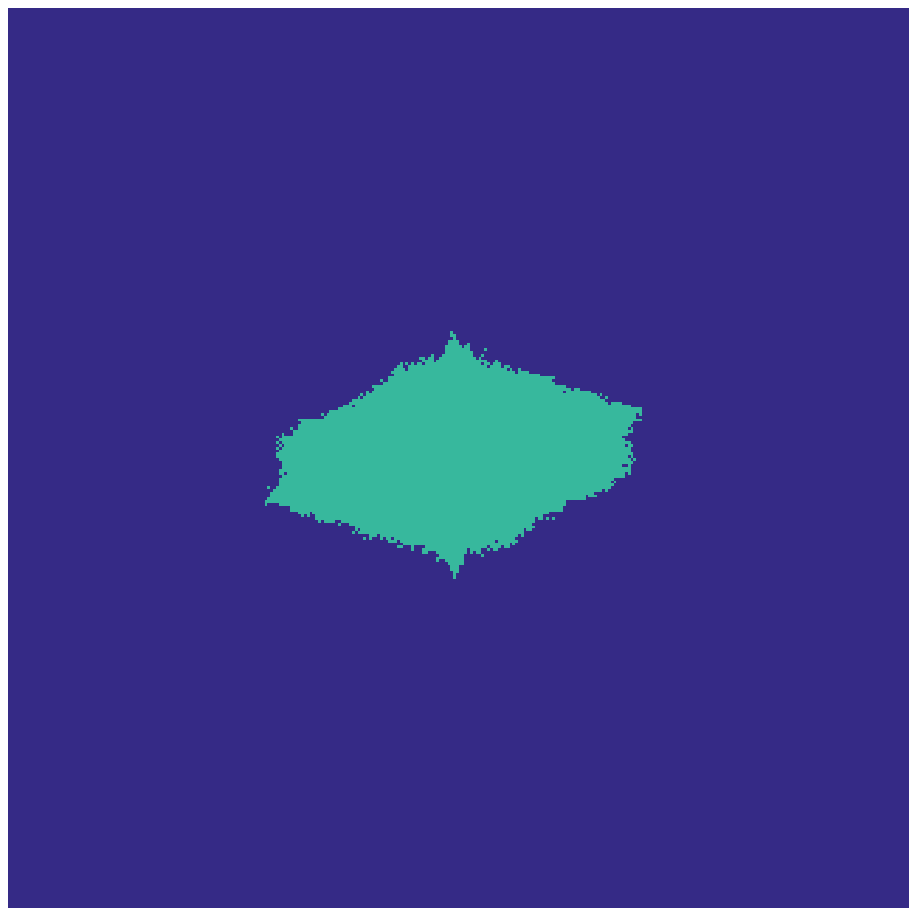} &
\hspace{-4mm}\includegraphics[width=.3\columnwidth]{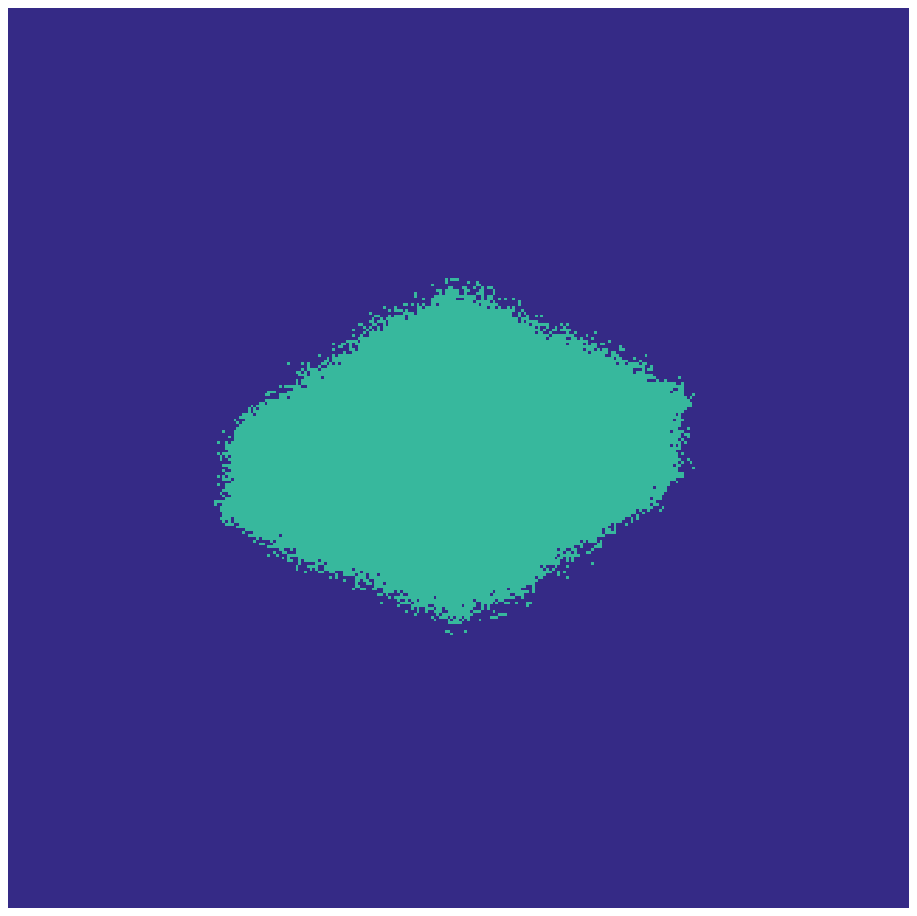} &
\hspace{-4mm}\includegraphics[width=.3\columnwidth]{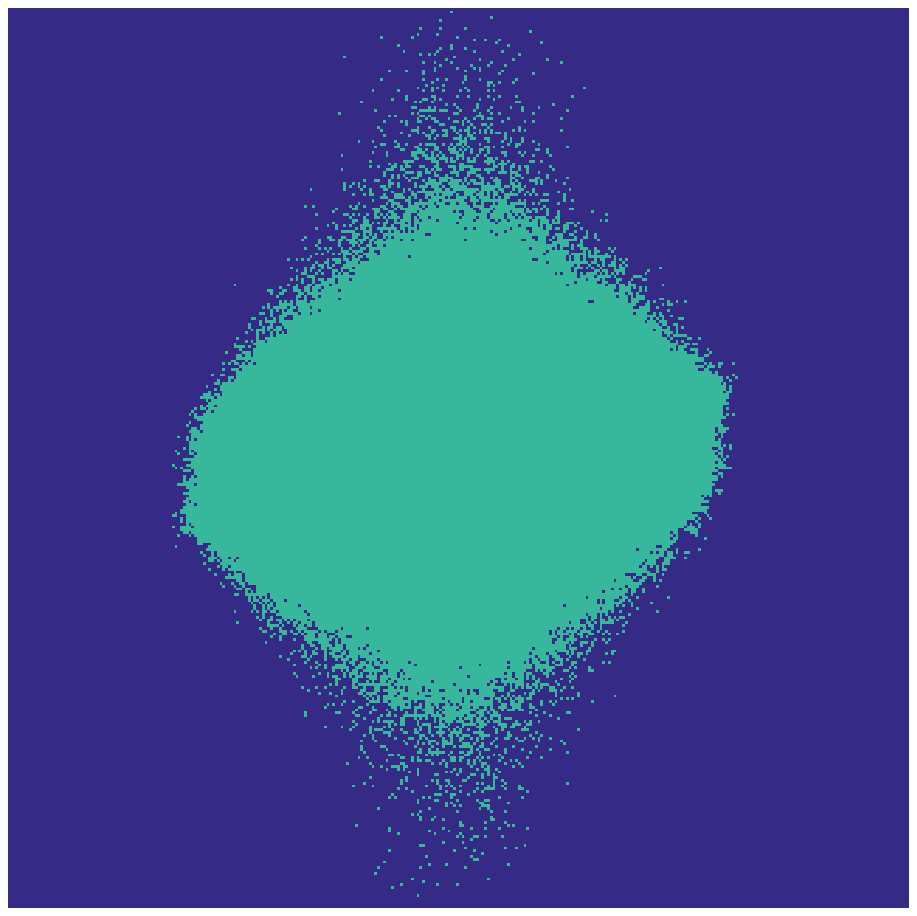} \\
 6.25\% sampling & 12.5\% sampling & 25\% sampling \\ 
\end{tabular}
\caption{First row: the subsampling maps of the tuned random variable sampling scheme \cite{Lustig2007}. Second row: the maps given by our learning-based approach.}
\label{fig:MRI_maps}
\end{figure}

Figure \ref{fig:MRI_maps} illustrates the best performing randomized indices and our learned set of indices in the $x-y$ plane of the $k$-space. Both the variable density approach \cite{Lustig2007} and our learning-based approach concentrates its sampling budget on the low frequencies, however the latter is endowed with the capability to adapt its frequency selection to the frequency content of the training signals instead of assuming a circularly symmetric selection.
%
%
%
%

\begin{table}[!h]
\caption{Average PSNR on the test data}\label{tab:mri}
\centering
\vspace{5pt}
\begin{tabular}{|c|c|c|c|}	
\hline
\multirow{2}{*}{Indices} 	& \multicolumn{3}{c|}{Sampling rate} 		\\ \cline{2-4}
 					& $6.25\%$	& $12.50\%$	& $25\%$ \\ 
\hline
						
Best-n approx.			& $25.29$ dB	& $26.36$ dB        & $28.35$ dB\\ \hline

Lustig \textit{et al.} 	& $24.51$ dB	& $25.11$ dB 		& $26.05$ dB \\ \hline
$\text{This work}$ 		& $24.66$ dB 	& $25.18$ dB 		& $26.12$ dB 	\\
\hline
\end{tabular}

\end{table}

\tabcolsep=0pt  

\begin{figure}[htb]
\centering
\begin{tabular}{cccc}
\includegraphics[width=.11\textwidth]{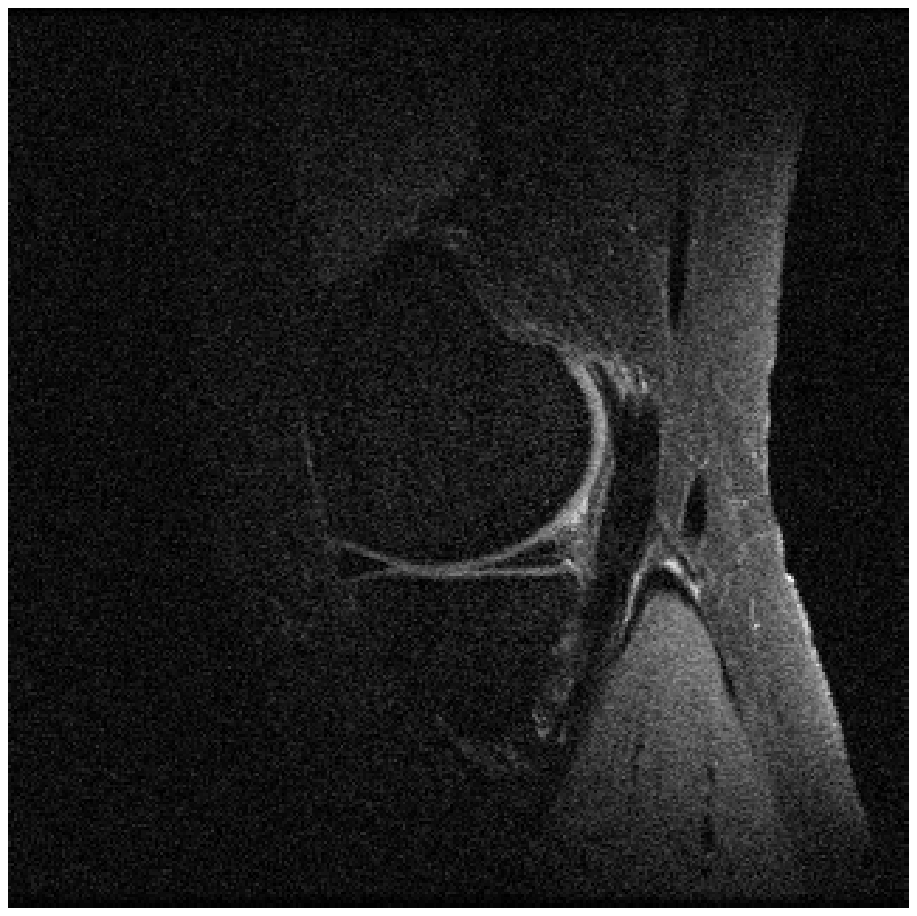} &
\hspace{+1mm}\includegraphics[width=.11\textwidth]{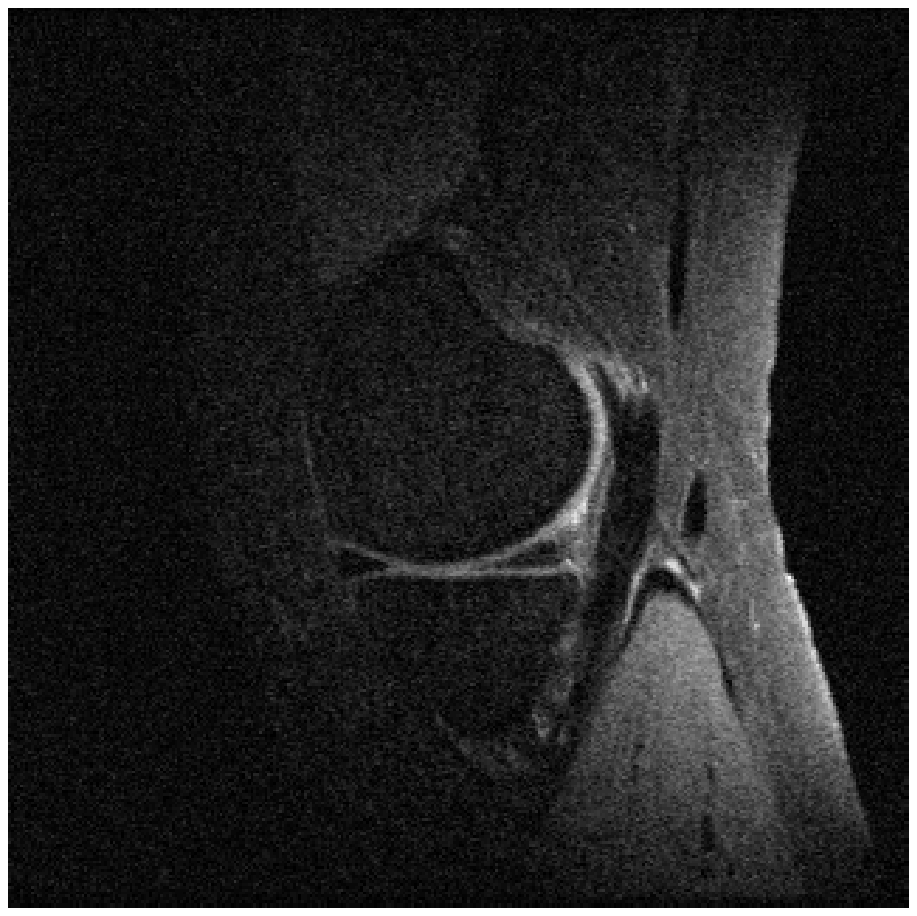} &
\hspace{+1mm}\includegraphics[width=.11\textwidth]{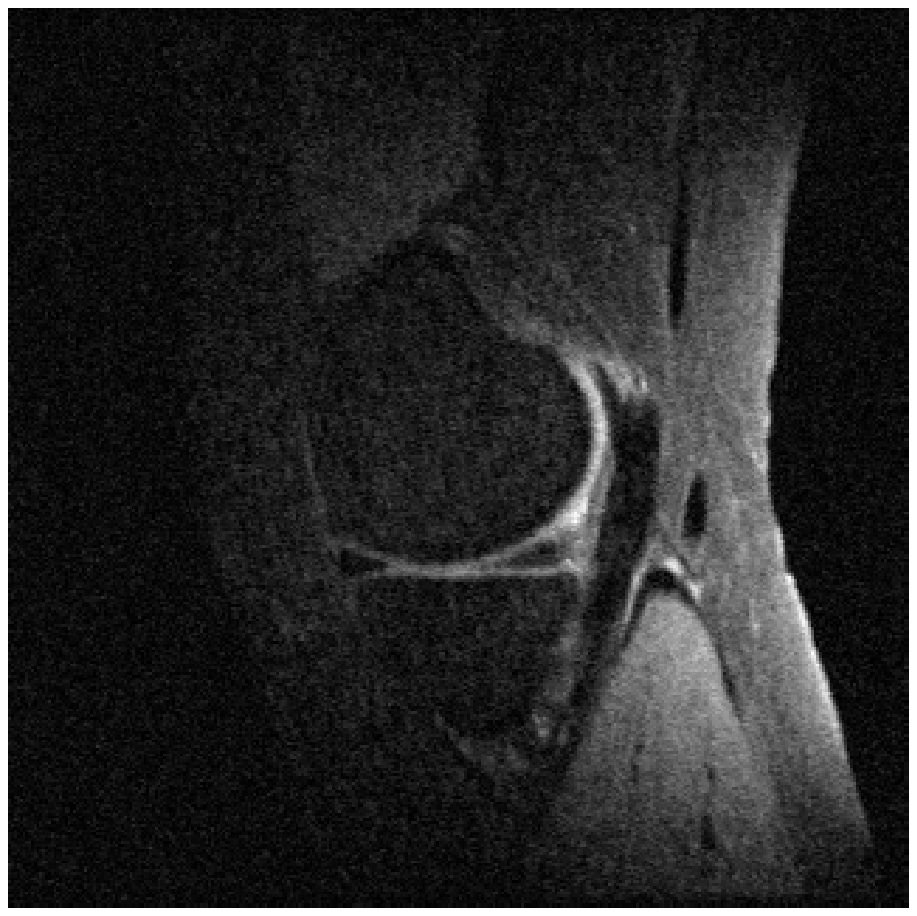} &
\hspace{+1mm}\includegraphics[width=.11\textwidth]{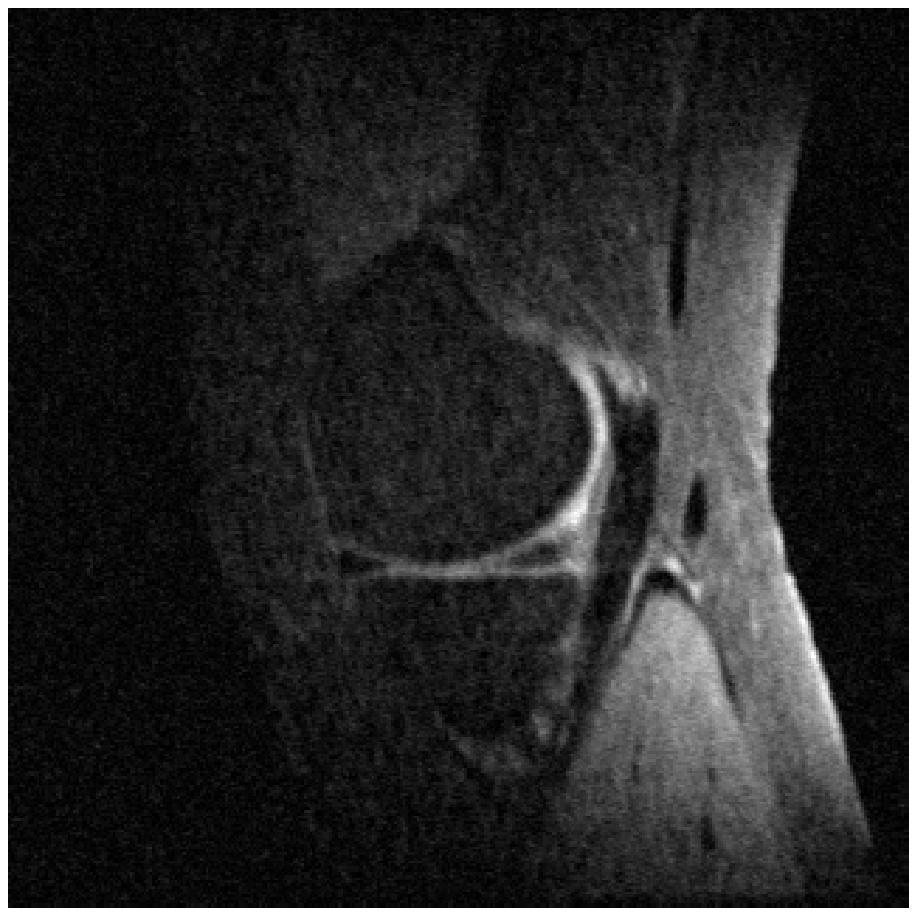} \\ [0mm]

\text{\footnotesize{${\text{learning-based}}$}} & 
\hspace{1mm}\includegraphics[width=.11\textwidth]{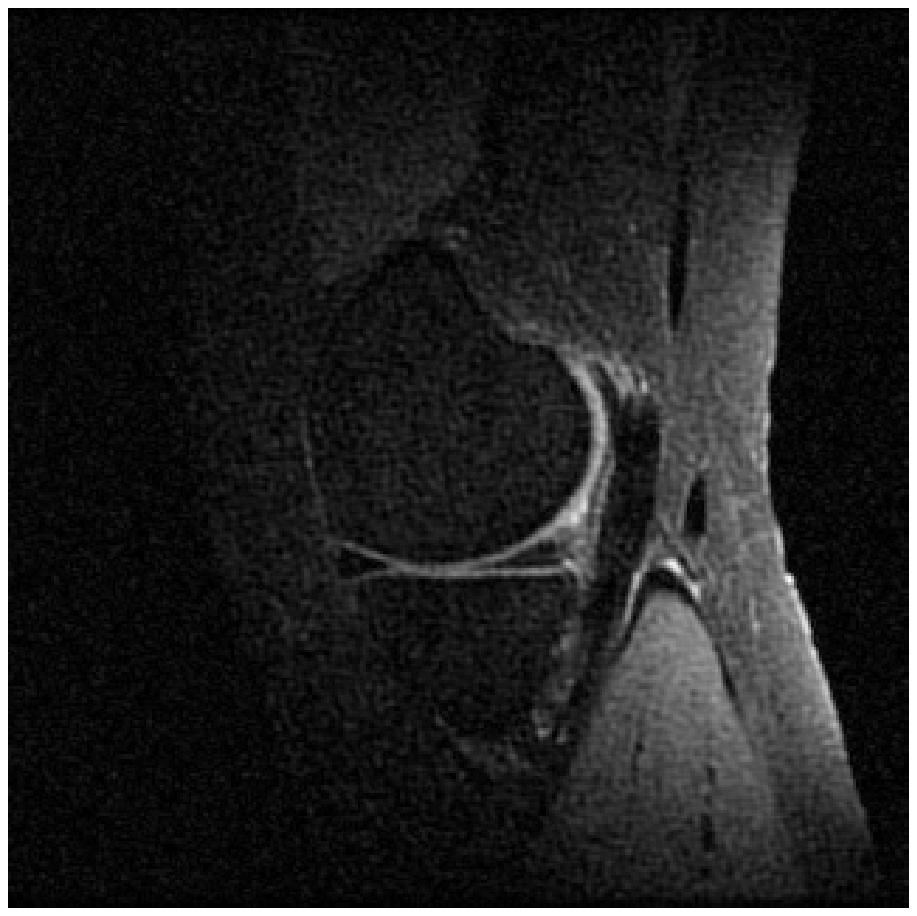} &
\hspace{1mm}\includegraphics[width=.11\textwidth]{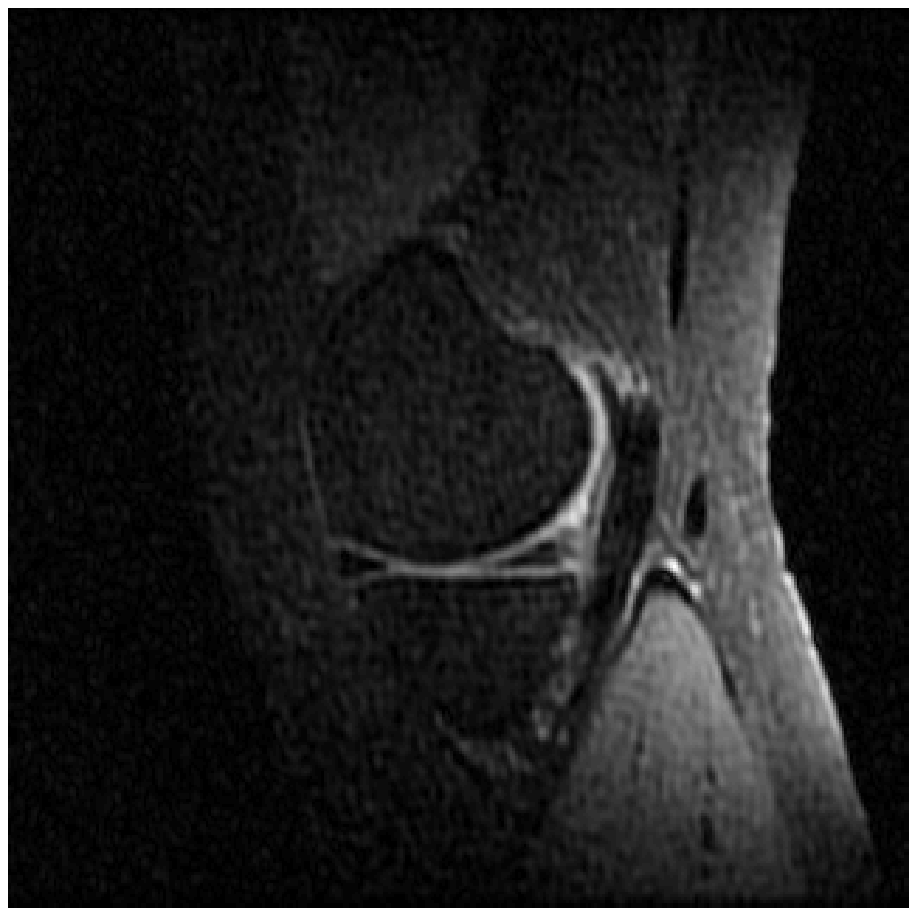} &
\hspace{1mm}\includegraphics[width=.11\textwidth]{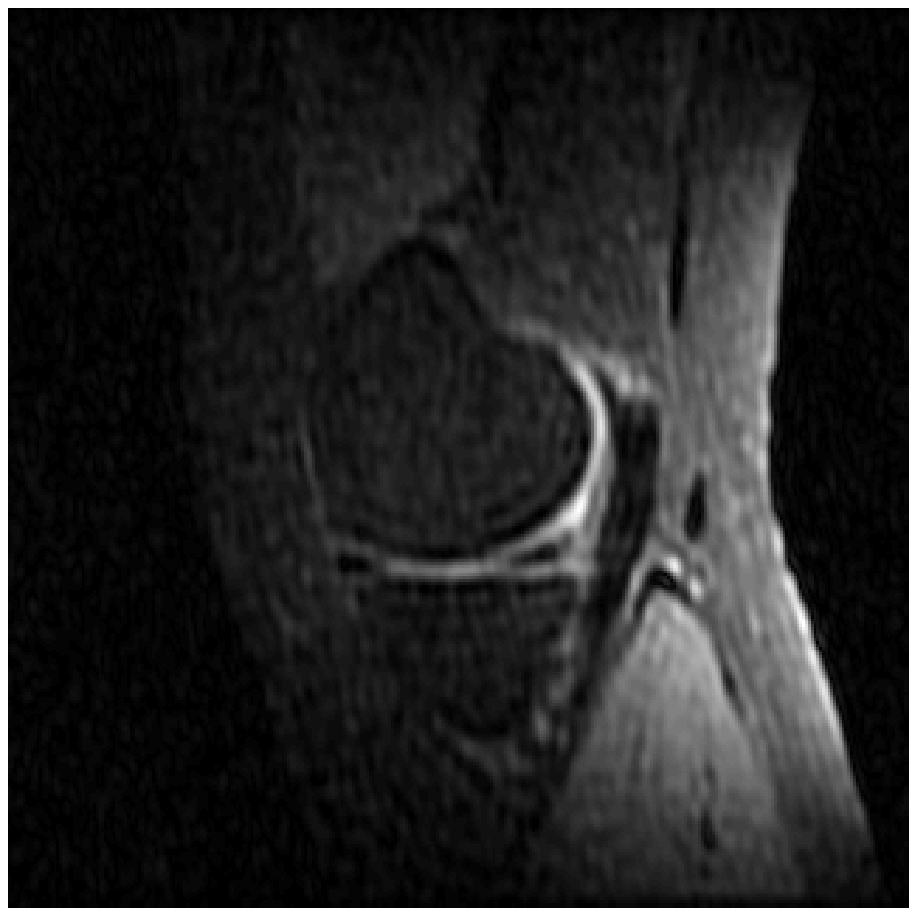} \\ 

\text{\footnotesize{Lustig \textit{et al.}}} & 
\hspace{1mm}\includegraphics[width=.11\textwidth]{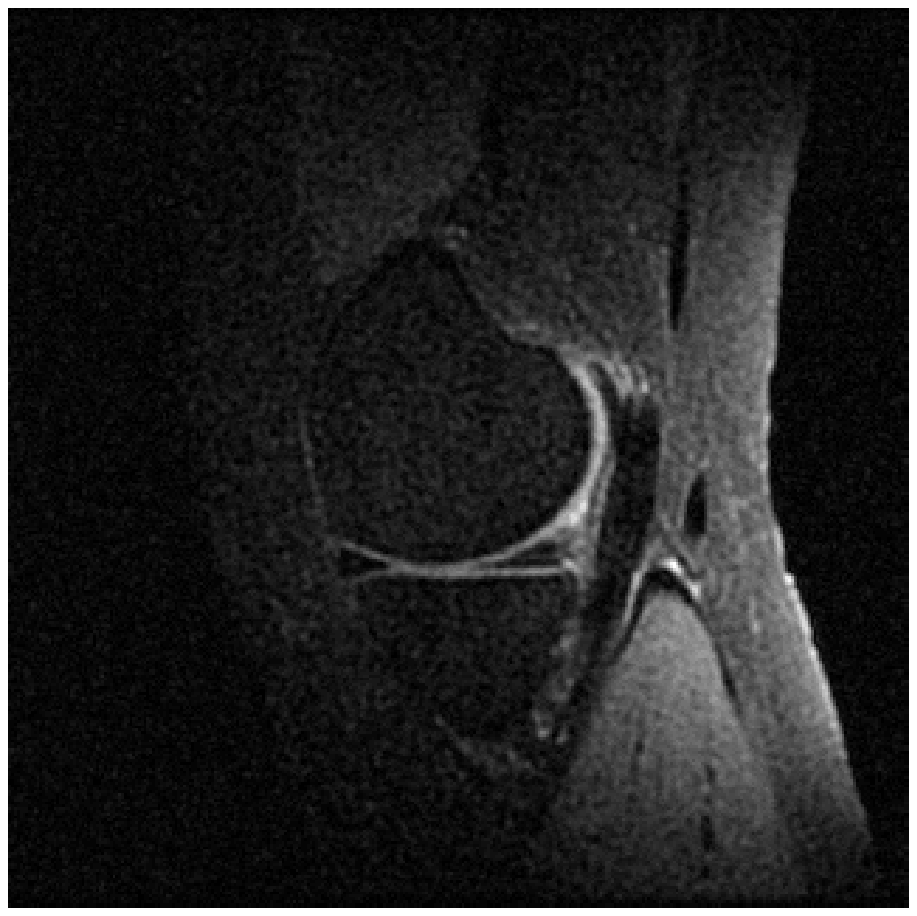} &
\hspace{1mm}\includegraphics[width=.11\textwidth]{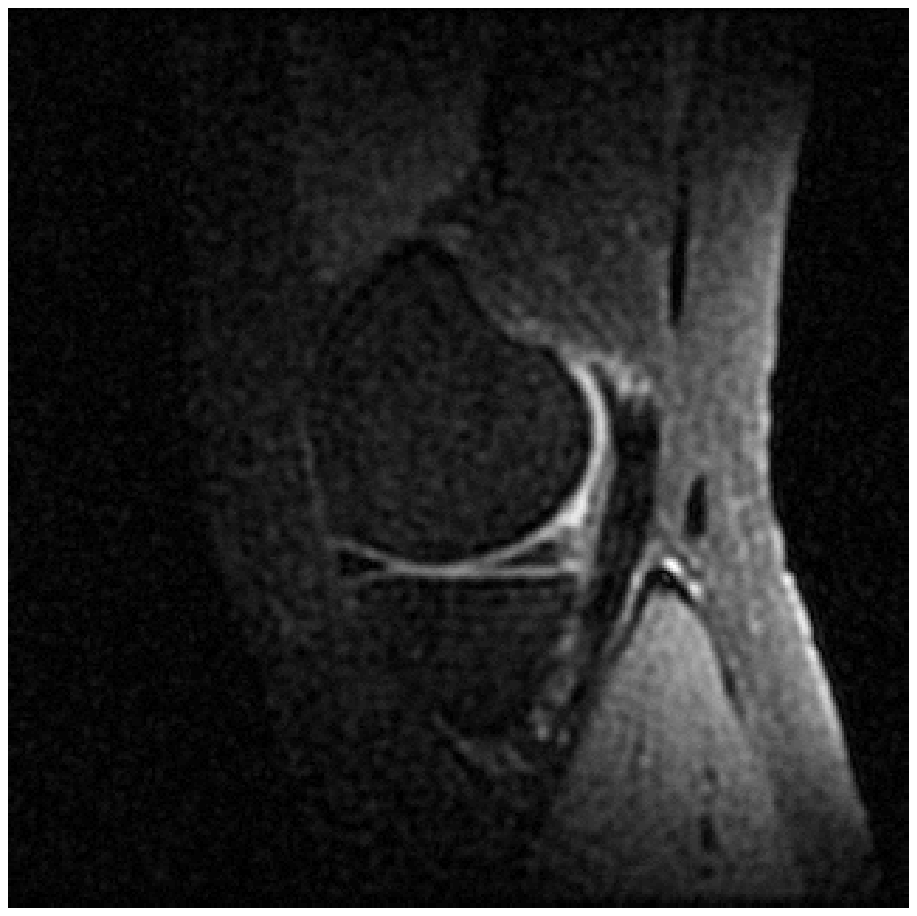} &
\hspace{1mm}\includegraphics[width=.11\textwidth]{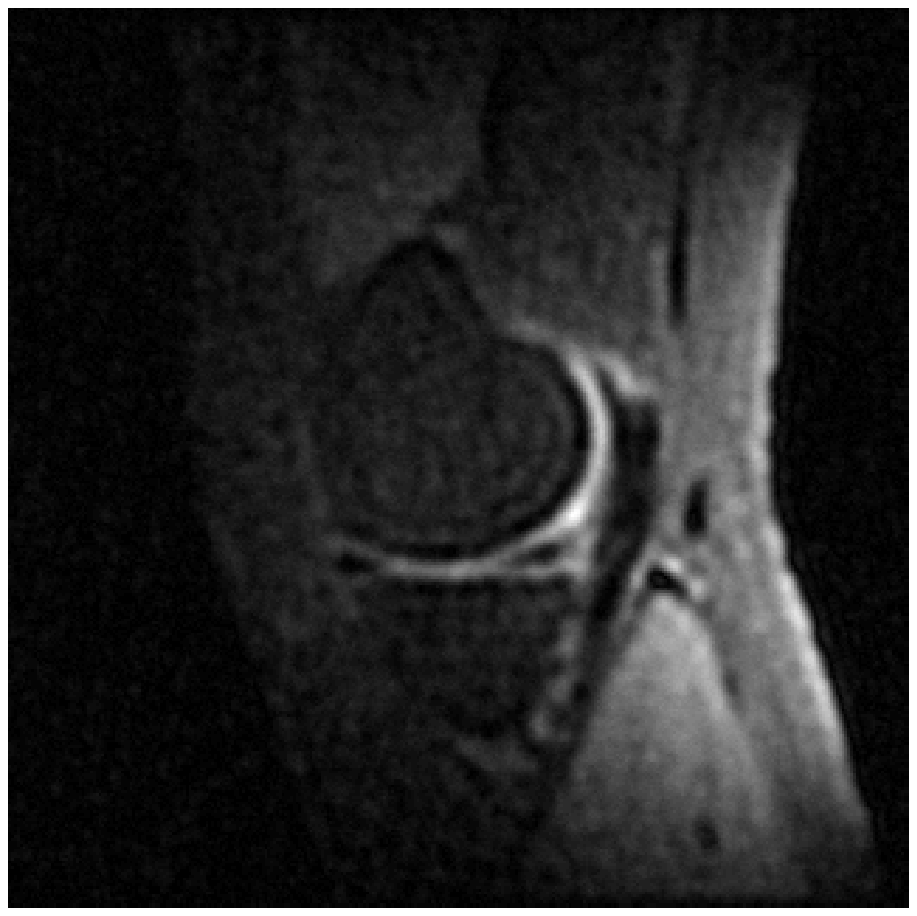} \\ [0mm]
& 25\%  & 12.5\%  & 6.25\%  \\ 
\end{tabular}
\caption{MRI reconstructions of both schemes at different subsampling rates for a knee slice of patient \#13, whose fully sampled reconstruction is shown on the top left. }
\label{fig:MRI_recon}
\end{figure}

Table \ref{tab:mri} shows the performance of both approaches on the test data, in addition to the error lower-bounds obtained by the best $n$-sample approximations with respect to the Fourier basis. It appears that the learning based approach slightly outperforms the randomized variable density based approach.
  
However, the slight numerical improvements are actually accentuated when we look at the details of reconstructions, shown in Figure \ref{fig:MRI_recon}  for the test Patient \#13. It is clear that the learning-based reconstructions provide more details especially for $6.25\%$ and $12.5\%$. 



\section{Discussions}

The essential idea of the learning-based approach can be summarized as follows: Fix a decoder, and find the optimal sub-sampling pattern that minimizes the corresponding expected recovery error, which can be approximated by empirical risk minimization. The performance is essentially determined by the distribution of signal ensemble.

In this paper, we consider the linear decoder for computational efficiency, and it works well on the ensemble of MRI images. For other signal ensembles, it is possible to have a better recovery error performance by a non-linear decoder, such as basis pursuit or the Lasso, and realize a trade-off between computational complexity and recovery performance. Note that the idea of the learning-based approach still applies, while the empirical risk minimization formulation for choosing the sub-sampling pattern should be modified accordingly given the decoder. We are currently working in this research direction.

\section{Acknowledgement}
The authors would like to thank Baran G\"{o}zc\"{u} and Luca Baldassarre for providing numerical results, and Jonathan Scarlett for the complexity discussion.

\section{Proofs}

\subsection{Proof of (\eqR)} \label{sec_prof_eqR}

In fact, the equality holds deterministically, as
\begin{align}
&\norm{ \hat{x}_{\Omega} - \xtrue }_2^2 \notag \\
&\quad = \norm{ \hat{x}_{\Omega} }_2^2 - 2 \left\langle \hat{x}_{\Omega}, \xtrue \right\rangle + \norm{ \xtrue }_2^2 \notag \\
&\quad = \norm{ \mathcal{F}^H P_{\Omega}^T P_{\Omega} \mathcal{F} \xtrue }_2^2 - 2 \left\langle F^H P_{\Omega}^T P_{\Omega} \mathcal{F} \xtrue, \xtrue \right\rangle + \norm{ \xtrue }_2^2 \notag \\
&\quad = \norm{ P_{\Omega} \mathcal{F} \xtrue }_2^2 - 2 \norm{ P_{\Omega} \mathcal{F} \xtrue }_2^2 + \norm{ \xtrue }_2^2. \notag
\end{align}
In the third equality, we used the fact that $A A^\dagger A = A$ for any matrix $A$ and its Moore-Penrose generalized inverse $A^\dagger$, by setting $A := P_{\Omega} \mathcal{F}$.

\subsection{Proof of Proposition \propMain}
\vspace{-3mm}
It suffices to choose $\varepsilon_m$ such that with probability at least $1 - \beta$,
\begin{equation}
\Delta_m := \mathbb{E}\, f_{\Omega_{\text{opt}}} ( x ) - \mathbb{E}\, f_{\Omega_m} ( x ) \leq \varepsilon_m. \notag
\end{equation}

We note that
\begin{align}
\Delta_m = \, \ & \left( \mathbb{E}\, f_{\Omega_{\text{opt}}} ( x ) - \Eemp \, f_{\Omega_{\text{opt}}} ( x ) \right) + \notag \\
& \left( \Eemp \, f_{\Omega_{\text{opt}}} ( x ) - \Eemp \, f_{ \Omega_m } ( x ) \right) + \notag \\
& \left( \Eemp \, f_{ \Omega_m } ( x ) - \mathbb{E}\, f_{\Omega_m} ( x ) \right). \notag
\end{align}
The second summand on the right-hand side cannot be positive by definition. Then we have
\begin{equation}
\Delta_m \leq 2 \max_{\Omega} \set{ \abs{ \Eemp \, f_{\Omega} ( x ) - \mathbb{E}\, f_{\Omega} ( x ) } : \Omega \in \mathcal{A} } \notag
\end{equation}
where $\mathcal{A} := \set{ \Omega : \Omega \subset \set{ 1, \ldots, p }, \abs{ \Omega } = n }$. 

As the random variables $f_{\Omega} ( x )$ are bounded ($0 \leq f_{\Omega} ( x ) \leq 1$), we can use Hoeffding's inequality and the union bound to obtain an upper bound of $\Delta_m$ that holds with high probability, as in \cite[B.3]{Audibert2007}.




\bibliographystyle{IEEEtranS}
\bibliography{list}

\end{document}